\documentclass[preprint]{aastex}
\usepackage{emulateapj5}
\shortauthors{Vadawale et al.}
\shorttitle{Various type of radio emissions in GRS~1915+105}
\slugcomment{Accepted in The Astrophysical Journal}

\begin{document}

\title{On the origin of the various types of radio emission in 
GRS~1915+105 }

\author{S. V. Vadawale\altaffilmark{1}, 
        A. R. Rao\altaffilmark{1},
        S. Naik\altaffilmark{1 \huge{*}},
        J. S. Yadav\altaffilmark{1},
	C. H. Ishwara-Chandra\altaffilmark{2},
	A. Pramesh Rao\altaffilmark{2}
	and G. G. Pooley\altaffilmark{3}}

\altaffiltext{1}
{Tata Institute of Fundamental Research, 
Homi Bhabha Road, Mumbai 400 005, India\\
E-mail: santoshv@tifr.res.in}
\altaffiltext{2}
{National Center for Radio Astrophysics, Post Bag No. 3,
Ganeshkhind, Pune 411 007, India}
\altaffiltext{3}
{Mullard Radio Astronomy Observatory, Cavendish Laboratory,
Madingley Road, Cambridge CB3 0HE\\
~~~~~~~~~~$^*$ : Present Address : Department of Physics, University College Cork, Cork, Ireland}

\begin {abstract}

We investigate the association between the radio ``plateau'' states and 
the large superluminal flares in GRS~1915+105 and propose a qualitative 
scenario to explain this association. To investigate the properties of 
the source during a superluminal flare, we present GMRT observations 
during a radio flare which turned out to be a pre-plateau flare as shown 
by the contemporaneous Ryle telescope observations. A major superluminal 
ejection was observed at the end of this ``plateau'' state (Dhawan et al.
2003), associated with highly variable X-ray emission showing X-ray soft 
dips. This episode, thus has all the three types of radio emission: a  
pre-plateau flare, a ``plateau'' state and superluminal jets. We analyze 
all the available RXTE-PCA data during this episode and show that: 
(1) the pre-flare ``plateau'' state consists of a three-component X-ray 
spectra which includes a multicolor disk-blackbody, a Comptonized 
component and a power-law and (2) the Compton cloud, which is responsible 
for the Comptonizing component, is ejected away during the X-ray soft dips. 
We investigate all the available monitoring data on this source and 
identify several candidate superluminal flare events and analyze the 
contemporaneous RXTE pointed observations. We detect a strong correlation 
between the average X-ray flux during the ``plateau'' state and the total 
energy emitted in radio during the subsequent radio flare. We find that 
the sequence of events is similar for all large radio flares with a fast 
rise and exponential decay morphology. Based on these results, we propose 
a qualitative scenario in which the separating ejecta during the superluminal 
flares are observed due to the interaction of the matter blob ejected during 
the X-ray soft dips, with the steady jet already established during the 
``plateau'' state. This picture can explain all types of radio emission 
observed from this source in terms of its X-ray emission characteristics.
\end {abstract}


\section{Introduction}

The black-hole candidate GRS~1915+105 is one of the most studied X-ray 
sources in our Galaxy due to its remarkably rich phenomenological 
diversity across the entire electromagnetic spectrum except the visible 
band. It was first detected in 1992 with the WATCH instrument on-board 
the GRANAT satellite (Castro-Tirado, Brandt, \& Lund 1992). In contrast 
with other transient X-ray sources, it has never been switched off since 
detection. Its optical counter-part has not been detected due to a  very 
high line-of-sight absorption ($\sim$26 magnitude, Chaty et al. 1996). 
In the absence of optical observations, the binary parameters as well as 
the nature of the companion were unknown till 2001. Finally nine years 
after its detection, Greiner et al. (2001), by near-infrared spectroscopic 
observations with VLT, established it as a Low Mass X-ray Binary (LMXB) 
system with a K-M III type companion star of   mass 1.2$\pm$0.2 M$_{\odot}$.
The black hole mass was determined to be 14$\pm$4 M$_{\odot}$ which is the 
largest mass known so far for any black hole candidate in an X-ray binary. 
GRS~1915+105 achieved the status of micro-quasar much earlier in 1994 when  
Mirabel \& Rodriguez (1994) detected episodes of superluminal ejections 
from this source, first time in any Galactic source. Observations of many 
such episodes of superluminal ejections during the past eight years make 
GRS~1915+105 distinct among other micro-quasars in our Galaxy. Apart from 
huge superluminal radio flares, GRS~1915+105 also shows various other 
types of radio emission like radio oscillations (Pooley \& Fender 1997) 
and extended periods of steady, flat spectrum radio emission known as 
radio ``plateau'' states (Fender et al. 1999). These radio ``plateau'' 
states are particularly interesting due to their association with the 
large radio flares. It is found that the ``plateau''  states are almost 
always followed by large radio flares with steep spectrum (Hannikainen 
et al. 1998; Fender et al. 1999; Klein-Wolt et al. 2002), though so far 
there is  no satisfactory explanation for this observed association.

The true glory of this source came into notice after the launch of NASA's
{\em Rossi X-ray Timing Explorer (RXTE)} in 1996. Pointed observations 
with RXTE-PCA showed extremely rich and exciting morphology in the X-ray 
intensity patterns (Greiner, Morgan, \& Remillard 1996; Morgan, Remillard, 
\& Greiner 1997; Muno, Morgan, \& Remillard 1999). Similar diversity in 
the X-ray emission from this source was also observed with the {\em Indian 
X-ray Astronomy Experiment (IXAE)} (Paul et al. 1997, 1998; Yadav et al. 
1999). Belloni et al. (2000) carried out a detailed analysis of all the 
RXTE pointed observations in 1996-97 and classified the complex diversity 
of the  X-ray emission exhibited by this source into 12 classes on the 
basis of light curves and hardness ratios (one more variability class was 
identified later on - see Klein-Wolt et al. 2002; Naik, Rao \& Chakrabarti 
2002). They also put an important step forward in understanding the 
complexity of this source by showing that all the variability classes 
occur because of repeated transitions of the source in three basic states. 
They showed that the transitions between these three states, named as 
states A, B, and C, can be very fast i.e. within a few seconds. The states 
A and B are characterized by  a  soft spectrum with the total luminosity 
being low and high,  respectively. The state C is characterized by a hard 
spectrum and the presence of strong  Quasi Periodic Oscillations (QPO) in 
the power density spectrum.

	This source is probably the best example for a strong connection 
between the accretion disk, manifested in the X-ray band, and the jet, 
manifested in the radio/IR band, in a black hole system (Pooley \& Fender 
1997; Eikenberry et al. 1998; 2000; Mirabel et al. 1998). Study of the 
correlated behavior of the source in the X-ray and the radio/IR wave 
bands plays a very important role in the endeavor of understanding this 
source and many attempts have been made so far to relate the X-ray 
emission characteristics with the episodes of various types of radio 
emission. Muno et al. (2001) studied representative observations in 
detail and showed that the properties of the QPO, like frequency, 
phase-lag, coherence etc. are correlated with the radio emission. 
Rau \& Greiner (2003) showed that the properties of accretion disk 
are not correlated with the radio emission.  Naik \& Rao (2000) showed 
that the radio emission is high only during three variability classes: 
$\beta$, $\theta$ and a sub-class of $\chi$ ($\chi_1$/$\chi_3$ or 
$\chi_{RL}$.)\footnote { Belloni et al. (2000) divided class 
$\chi$ into four sub-classes $\chi1$, $\chi2$, $\chi3$ and $\chi4$ based 
on their order of occurrence. Out of these four sub-classes only $\chi_1$ 
and $\chi_3$ show high radio emission. However, later more such occurrences 
of class $\chi$ have been observed during which the radio emission is high. 
Thus, instead of numbered sub-classes, class $\chi$ can be divided into two
generic sub-classes radio-quiet ($\chi_{RQ}$) and radio-loud ($\chi_{RL}$).
The sub-class $\chi_{RL}$ corresponds to the radio ``plateau'' states.} 
They suggested that the radio emission is associated with the soft X-ray 
dips (short periods of state A) observed in the classes $\beta$ and $\theta$, 
apart from the sub-class $\chi_{RL}$. Naik et al. (2001) detected similar 
dips in the X-ray light curve during the presence of a large radio flare 
indicating their association with large radio flares.

Klein-Wolt et al. (2002) studied all simultaneous observations of the 
source by RXTE and Ryle Telescope and showed that the radio emission is 
associated with only state C and not with states A or B. They also found 
an one-to-one relation between series of long ($>$ 100 s), well separated 
state C intervals and radio oscillation events. They suggested a scenario, 
according to which the long uninterrupted state C intervals are associated 
with continuous jets which give ``plateau'' radio emission, whereas if 
the state C intervals are well separated by other states then the radio 
oscillation events are observed. A different picture, however, emerges 
when the association of the radio emission with the different sub-classes 
of $\chi$ (state C) are examined in detail (see discussion in 
section~\ref{scenario} for more details on this).

 	The sub-classes of class $\chi$ can be divided into two different
types of hard states: radio-loud hard state (i.e. $C_{RL}$) and radio-quiet
hard state (i.e. $C_{RQ}$) (Vadawale et al. 2001b; Trudolyubov 2001). By a
detailed spectral study of RXTE observations of the source belonging to
the two types of hard states, Vadawale et al. (2001b) showed that the 
wide band X-ray spectrum of state $C_{RQ}$ consists of two components: 
a multicolor disk-blackbody and a Comptonized component whereas the wide 
band X-ray spectrum of state $C_{RL}$ consists of three components: a 
multicolor disk-blackbody, a Comptonized component and a power-low. 
They further showed that the additional power-law component can be 
modeled as due to synchrotron radiation in X-rays, coming from the base 
of a continuous jet which is present only during the $C_{RL}$ states. 
This three-component picture gets further support from other results such 
as: only the Comptonized component is responsible for the low-frequency 
QPO (Rao et al. 2000) and both the QPO as well as the Comptonized component 
are absent during the soft dips (short state A periods) observed in class 
$\theta$ (Vadawale et al. 2001a). It should be noted that the radio-loud 
hard state also occurs as a part of other variability classes like 
$\theta$ and $\beta$. Hence we will refer it generically as $C_{RL}$, 
whereas the particular sub-class of class $\chi$ during which the radio 
emission is high, will be referred as $\chi_{RL}$.

	In this paper we explore the association of the radio ``plateau'' 
states with large superluminal radio flares and try to envisage a scenario 
to explain this association as well as the various types of radio emission 
observed from this source. Since there are not many detailed studies of 
such episodes of superluminal jet emissions, in the first part of the paper 
we present a detailed study of a complete sequence of flare morphology 
described, sequentially,  by a small flare, ``plateau'' state, a disturbed 
accretion disk and a superluminal jet. We report the observations of the 
source with {\em Giant Meter-wave Radio Telescope (GMRT)} at 1.28 GHz when 
the source exhibits a flare which turned out to be a pre-plateau flare. 
We also present the results of a detailed analysis of all the pointed RXTE 
observations during this flare cycle which confirm the three-component
X-ray spectral description of the pre-flare ``plateau'' state. In order
to confirm the three-component description during all occurrences of the 
``plateau'' state, we further carry out similar analysis of representative 
pointed RXTE observations prior to all huge radio flares identified in 
the GBI monitoring data on this source. Finally we summarize the results 
and propose a phenomenological model for different types of radio emission 
from this source.

\section{Observations : 2001 June-July radio flare episode}
\subsection{GMRT observations}

\begin{figure*}
\centering
\vskip 22.5cm
\includegraphics{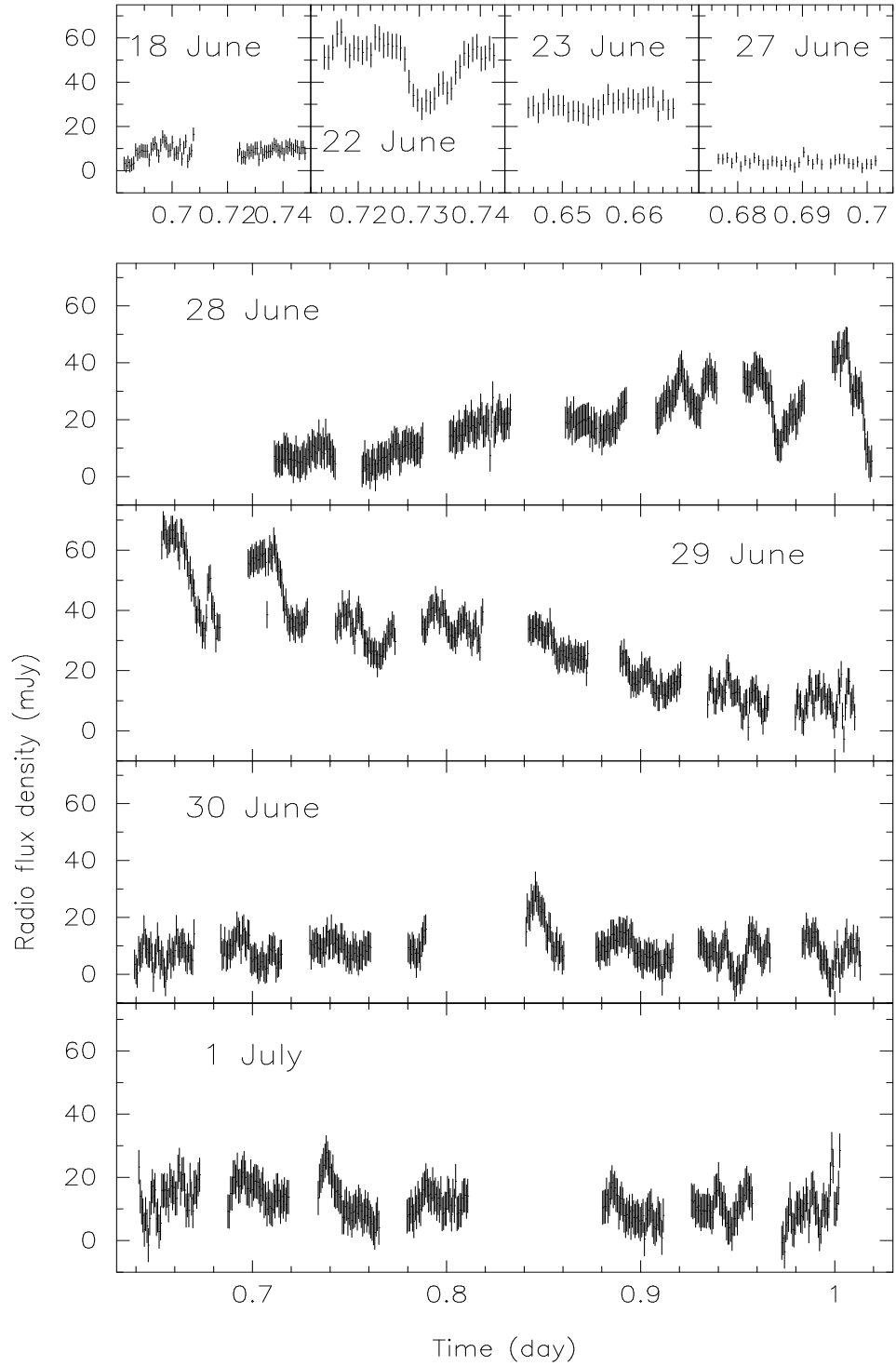}
\caption{Radio light curves of GRS~1915$+$105 at 1.28 GHz observed with 
the Giant Meter-wave Radio Telescope (GMRT) during 2001 June-July.}
\label{gmrt}
\end{figure*}

Radio observations of the micro-quasar GRS~1915+105, at 1.28 GHz with 
a band pass of 16 MHz, were carried out with GMRT (Swarup et al. 1991) 
on 2001 June 18, 22, 23, 27, 28, 29, 30 and July 1. The flux density 
scale was set by observing the primary calibrator 3C286 / 3C48. A phase 
calibrator was observed before and after a 45 minutes scan on GRS~1915+105. 
The integration time during the source observations was 32 s. The data 
reduction and analysis were done by using the standard Astronomical Image 
Processing System (AIPS) package. The observations on 2001 June 18, 22, 
23, and 27 were of $\sim$ 30-60 minutes, whereas on June 28, 29, 30 and 
July 1, the observations were of $\sim$ 8-10 hours. Light curves, with 
32 s integration time, were extracted from each of the individual 
observations of the source and are shown in Figure~\ref{gmrt}. 

It can be seen from the figure that the source was quite steady on June 
18 with a very low flux density ($\sim$10 mJy). The source was found 
to be bright in radio on June 22 with a flux density of $\sim$ 50 mJy,
along with a sharp dip with a decrease in the flux density by a factor 
of $\sim$ 2. On June 23 the source was steady at about 30 mJy. It was 
very weak ($\sim$ 5 mJy) on June 27. From June 28 onwards the observations 
were of long durations (8 -- 10 hours) and hence a wide range of flux 
variabilities could be studied. On June 28  the source was found to be 
very weak at the beginning of the observations (UTC 17 hour), but the 
radio flux density gradually increased to $\sim$ 50 mJy followed by a 
sharp decrease to $\sim$ 5 mJy at the end of observations (UTC 24 hour). 
On the other hand, on June 29, the source was bright in radio with flux 
density $\sim$ 70 mJy at the beginning which decayed down to 10 mJy 
towards the end. The source was relatively weak and steady on June 30 
and July 1. It can be seen that the radio emission is highly variable 
over time scales of a few tens of minutes, similar to that observed during
mini flares (or ``baby jets''). Analyzing many individual ``clean'' flares 
from the same data, Ishwara-Chandra et al. (2002) have tried to model these 
flares as due to adiabatically expanding synchrotron emitting clouds.

\subsection{Radio observations by other telescopes}

\begin{figure*}
\centering
\vskip 9.0cm
\includegraphics{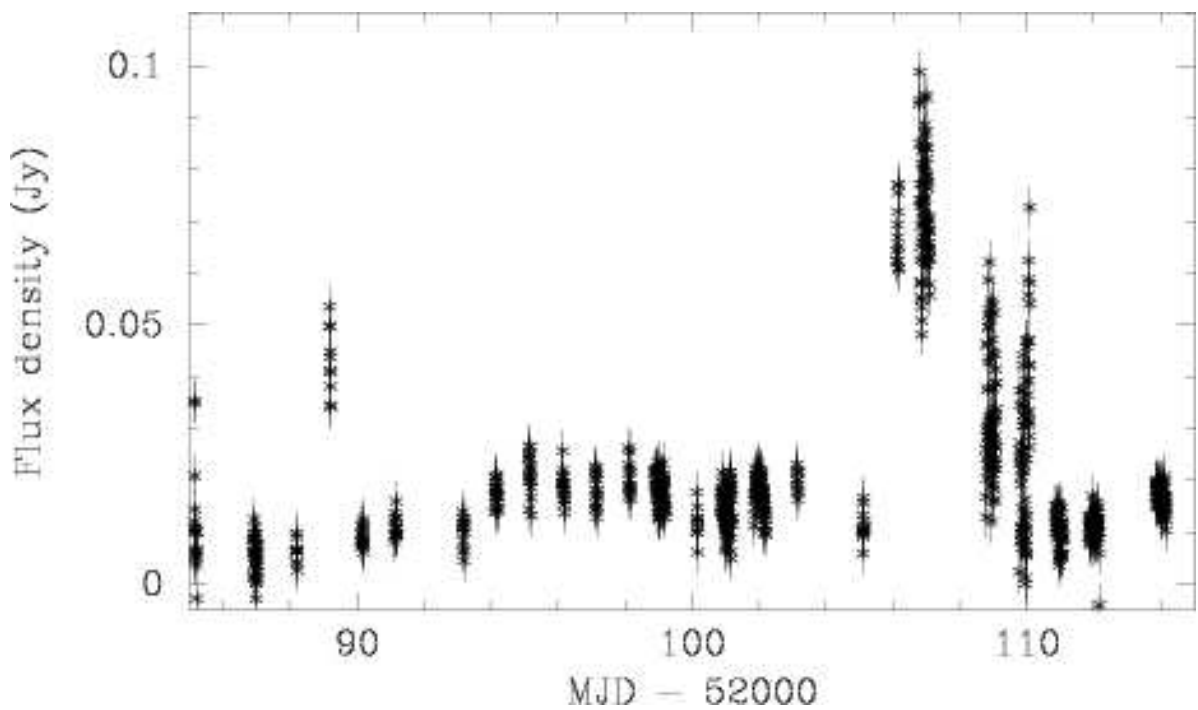}
\caption{Radio monitoring light curve of GRS~1915$+$105 at 15 GHz, obtained 
with the Ryle telescope. The presence of a large radio flare is seen on 2001 
July 16 (MJD 52106).}
\label{ryle}
\end{figure*}

To corroborate the results from  the GMRT observations, we obtained radio 
monitoring data of GRS~1915+105 from 2001 June 15 to July 25 observed with 
the Ryle telescope at 15.2 GHz (Pooley \& Fender 1997). Radio light curves 
at 15.2 GHz are shown in Figure~\ref{ryle}. A strong radio flare was 
observed on June 21 which was narrowly missed by GMRT. Radio flares with 
low peak flux density were also observed with both GMRT and Ryle telescopes 
on June 28 and 29. Similar flux densities at 1.4 GHz and 15.2 GHz suggest 
the flat nature of the spectrum during the flares. It can be seen, from 
Figure~\ref{ryle}, that the source went into a radio-loud ``plateau'' 
state after the GMRT observations which is followed by a radio flare on 
July 16, observed with the {\it RATAN} telescope (Dhawan et al. 2003). 
The flux density of this flare was much higher at lower frequencies (as 
observed by RATAN) compared to that at 15 GHz (as observed by Ryle) 
suggesting a steep spectrum during the flare. This source was also 
observed by VLBA on July 16, 17 and 18 at 2.25, 8.3 and 15 GHz. VLBA 
observations clearly showed an ejecta well separated from the core 
(Dhawan et al. 2003). 

These observations suggest that this superluminal flare episode followed
a typical sequence of events observed during the past episodes of 
superluminal flares. It started with a small flare, which was observed 
by GMRT, followed by a long radio ``plateau'' state, followed by 
superluminal ejections. It is thought that the different types of radio 
emission are results of changes in the accretion flow which are directly 
reflected in X-rays. Thus to understand the sequence of these events it 
is necessary to make a detailed analysis of the X-ray observations during 
this time period.

\subsection{RXTE pointed observations}

To investigate the X-ray properties of the source during and before 
this 2001 July 16 flaring episode, we analyzed all the 20 available 
RXTE pointed observations of GRS~1915$+$105 from 2001 June 30 to 2001 
July 16. Standard procedures for data reduction, response matrix 
generation, and background estimation were followed using the software 
package HEASoft 5.2. For spectral analysis we used Standard-2 mode data 
from Proportional Counter Array (PCA) and Archive mode data from HEXTE 
Cluster-0. We added the count rate spectrum of all layers of individual 
Proportional Counter Units (PCUs) and all available PCUs together. 
Analyzing the spectra of Crab with the new response matrices, it is found 
that 0.5\% systematic error is required to achieve an acceptable fit and 
the value of the power-law photon index is also found to be close to the 
canonical value of 2.1. Systematic error of 0.5\% is, therefore, added to 
the PCA spectra of all the observations used in the present work. 

\begin{figure*}
\centering
\vskip 14.0cm
\includegraphics{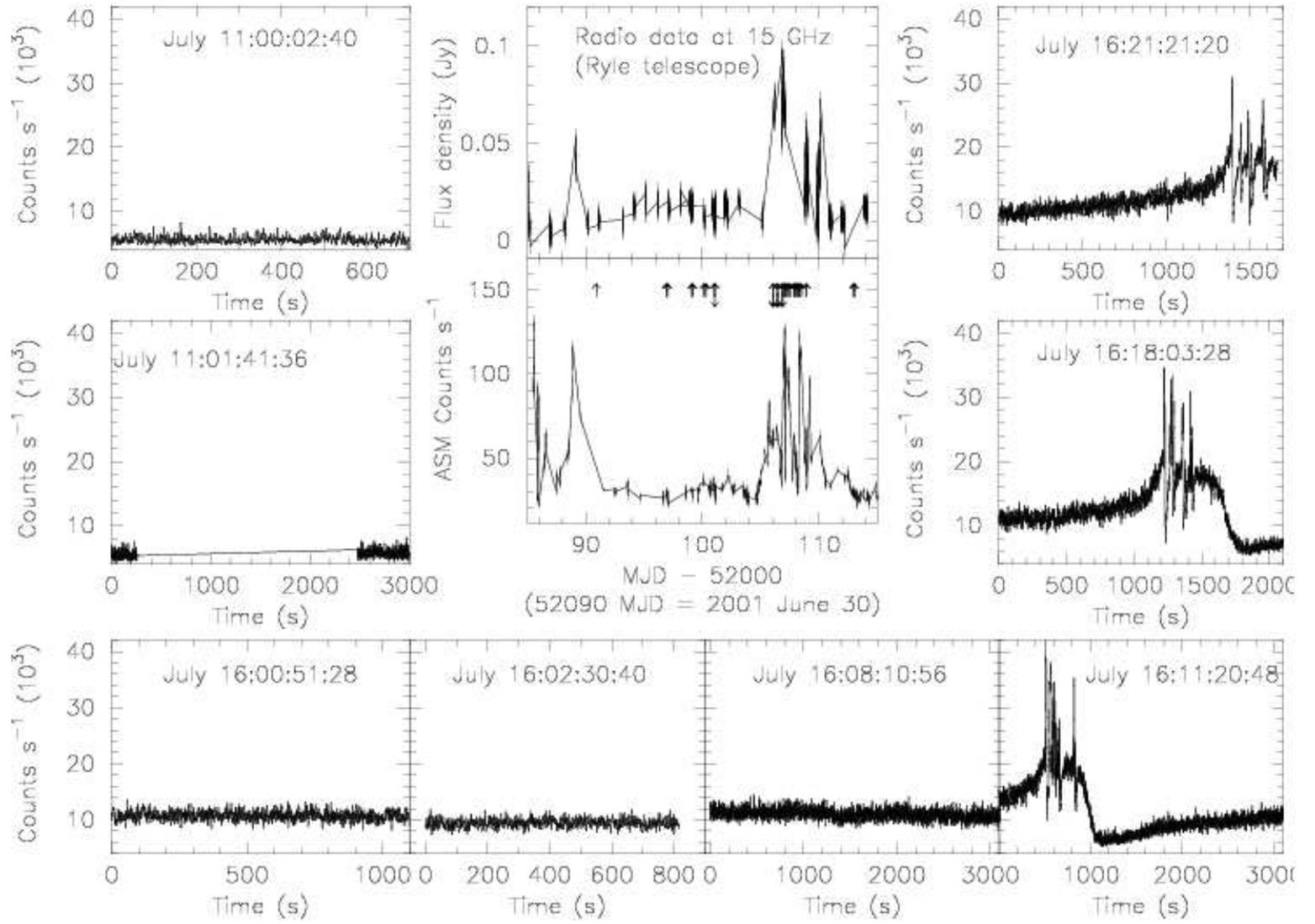}
\caption{2--60 keV RXTE-PCA light curves of GRS~1915$+$105 during and 
before the large radio flare on 2001 July 16. Central two panels show 
radio and X-ray monitoring light curves from 2001 June 25 to 2001 July 
25 obtained from Ryle telescope and RXTE/ASM. Arrows in upward direction 
in the ASM panel show all pointed RXTE observations during this period 
whereas the observations, light curves of which are shown in the surrounding 
panels (ordered anti-clockwise, starting from top-left), are indicated by 
arrows in downward direction.}
\label{jul_lc}
\end{figure*}

X-ray light curves with 1 s time resolution (Standard-1 mode) are 
generated for all the 20 observations. It is found that during all 
the RXTE/PCA observations before 2001 July 16, the source was in a 
hard state (i.e. the source state is $C_{RL}$ whereas the class is 
$\chi_{RL}$). Light curves of all the RXTE/PCA observations on July 
16 and the last two RXTE/PCA observations before July 16, along with 
the X-ray (RXTE/ASM) and radio (Ryle telescope) monitoring light 
curves, are shown in Figure~\ref{jul_lc}. It can be seen from 
Figure~\ref{jul_lc} that the last three pointed RXTE/PCA observations 
on July 16 clearly show evidence for a disturbed accretion disc. The 
first three pointed observations on July 16  (when the radio flare is 
already on) show enhanced X-ray emission, which could be either an 
enhanced $\chi_{RL}$ class or the steady part of the X-ray emission 
during a disturbed accretion disc. Because of the gaps between the 
pointed observations, it is not possible to precisely identify the 
time of occurrence of the state change from the X-ray light curves.
We carried out a detailed timing and spectral analysis of all the 20 
RXTE observations to examine the X-ray properties so that the time of 
occurrence of the state change could be better understood.

\subsubsection{Timing Analysis}

\begin{figure*}
\vskip 13.5cm
\includegraphics{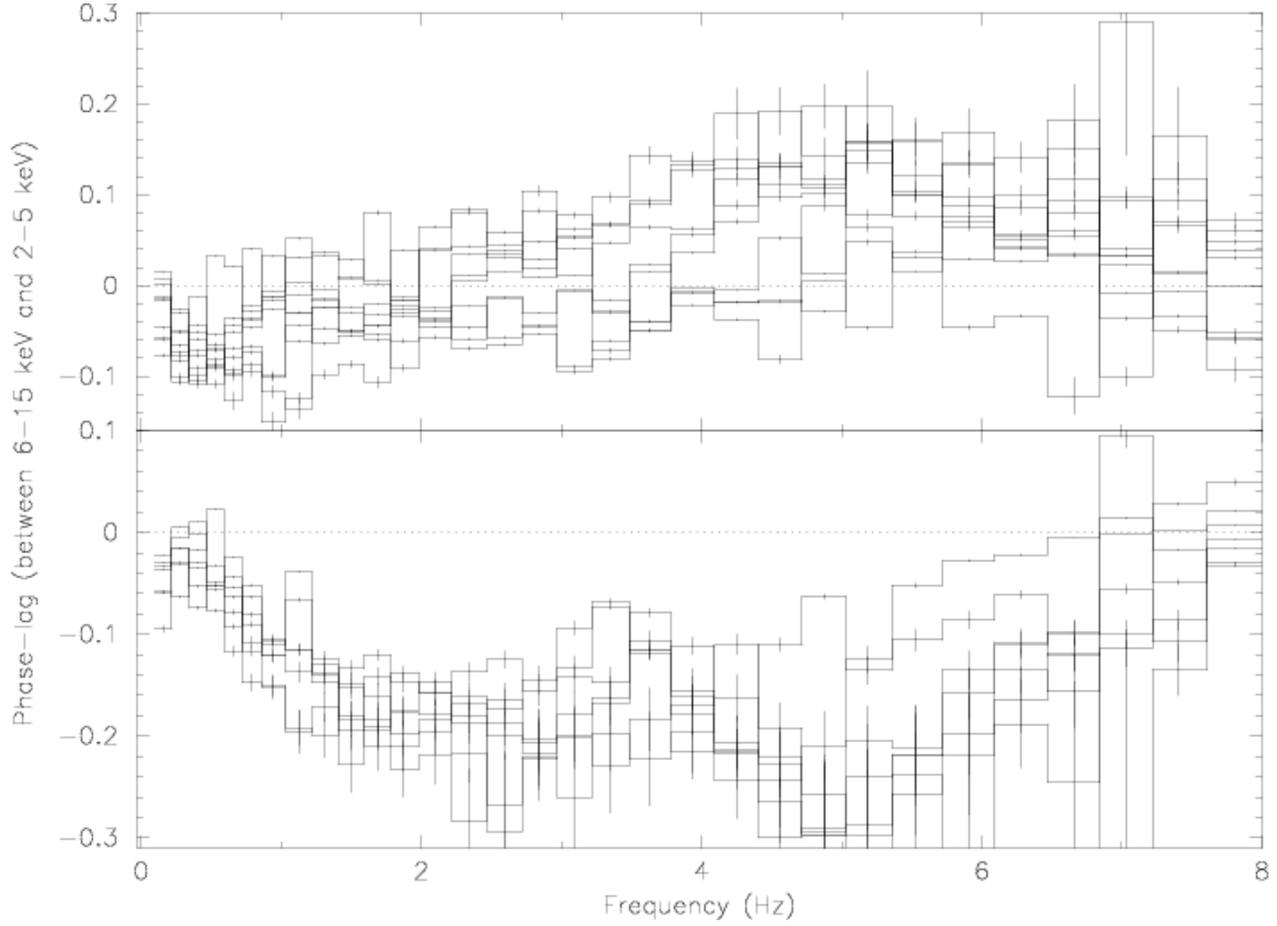}
\caption{The phase-lag spectra of all RXTE observations in the ``plateau'' 
state prior to the radio flare on 2001 July 16 (top panel) and observations 
during the onset of a radio flare on 2001 July 16 (bottom panel). Similar 
shape of the phase-lag spectra in each panel suggests that the physical 
conditions in the source are same during these periods. Distinct differences 
in the overall shape of the phase-lag spectra in both the panels suggests 
that the source conditions must have changed before 2001 July 16.}
\label{phase}
\end{figure*}

For timing analysis, Single-Bit mode data, available in the energy ranges 
of 3.6 -- 5.7 keV and 5.7 -- 14.8 keV, are used. Power density spectra (PDS) 
of 256 bin light curves are generated and co-added for every 16 seconds to
achieve better signal to noise ratio. Results of timing and spectral 
analysis are given in Table 1, along with the log of observations. The 
parameters for one representative observation per day (before the radio 
flare) and all observations carried out on July 16 (when the radio flare 
commenced) are given in the table. Quasi periodic oscillations (QPOs) in 
1 -- 3 Hz frequency range are seen in the PDS of all the RXTE observations 
made before July 16, whereas the PDS of all observations on July 16 show 
QPOs at $>$ 4 Hz (see Table~\ref{jul}a). Along with the PDS, phase-lag 
spectra are also generated by multiplying the imaginary part of the 
coefficients of the power density spectra in the above two energy ranges 
(Reig et al. 2000). Individual phase-lag spectra for each 16 s interval 
were also co-added. Phase-lag spectra of all the observations are shown 
in Figure~\ref{phase}. The top panel of Figure~\ref{phase} shows the 
phase-lag spectra of all observations made before July 16 (``plateau'' state) 
and the bottom panel shows the phase-lag spectra of all observations made 
on July 16 (including those showing flare/dips, however, these phase-lag 
spectra are calculated only for the steady portion of the light curve). 
It can be seen that there is a clear difference in the overall shape of 
the phase-lag spectra. In order to quantify the shape of the phase-lag 
spectra of the observations before and on July 16, we applied KS test 
which shows that these two distributions are different with 99.99\% 
confidence level significance. The reason for this difference in the 
shape of the phase-lag spectra is not clear, however, it is possible 
that the ejected matter during the X-ray disturbance (see next section) 
might be playing some role here as suggested by Pottschmidt et al. (2000).
However, the difference clearly shows that the source state before July 16 
is different from the state during the steady observations on July 16 and 
the disturbance in X-rays has already occurred before the first observation 
on July 16.

\subsubsection{Spectral Analysis}

XSPEC package (Ver. 11.2) is used for the spectral analysis. A simultaneous 
fit to the PCA (in the energy range of 3 - 35 keV) and HEXTE (in the energy 
range of 20 - 150 keV) data was carried out, keeping the relative 
normalization as a free parameter. A fit with any continuum model shows 
strong residuals near 6.4 keV and any fit without a Gaussian line at this 
energy is not acceptable. Such residuals, though with a low  amplitude, 
are also seen in the Crab spectra obtained with PCA. It is not clear what 
fraction of the residuals is really due to the source and what fraction 
is due to an instrumental artifact. The line feature at 6.4 keV, therefore, 
will not be discussed further in this paper. It is found that a model 
consisting of three components: a disk-blackbody, a power-law and a 
Comptonized component (CompTT in XSPEC, see Titarchuk 1994) is required 
for a statistically and physically acceptable fit to the source spectra 
(see Vadawale et al. 2001b for a comparison of the various models). The 
best-fit parameters along with the continuum flux of individual spectral 
components are given in Table~\ref{jul}. It can be seen from Table~\ref{jul} 
that the equivalent hydrogen column density is very high compared to the 
values reported in the literature (e.g. Muno et al. 1999; Vilhu et al. 2001). 
Spectral fitting  with this three-component model requires N$_H$ to be 
within 12 $-$ 15 $\times 10^{22} cm^{-2}$ with 99\% confidence level. 
The derived inner-disk radii (R$_{in}$) are also found to be high during 
these observations, compared to the reported values (e.g. Rau \& Greiner 
2003). However, these apparently high values of R$_{in}$ are physically 
plausible i.e. 15 -- 25 R$_g$ for a 14 M$_\odot$ black hole. The inner 
disk temperature (kT$_{in}$) and the plasma temperature (kT$_{e}$) for all 
the observations are $\sim$ 0.5 keV and 7--11 keV respectively and are 
acceptable. It should be noted that on June 30 (MJD 52090), when the radio 
emission was relatively low (the ``plateau'' started on July 1), the spectral 
and timing parameters are different from the rest of the observations with 
the value of N$_H$ being close to the reported value, which gives confidence 
for the applicability of this three-component model to the $C_{RL}$ state 
of this source.

\begin{figure*}
\vskip 18.9cm
\includegraphics{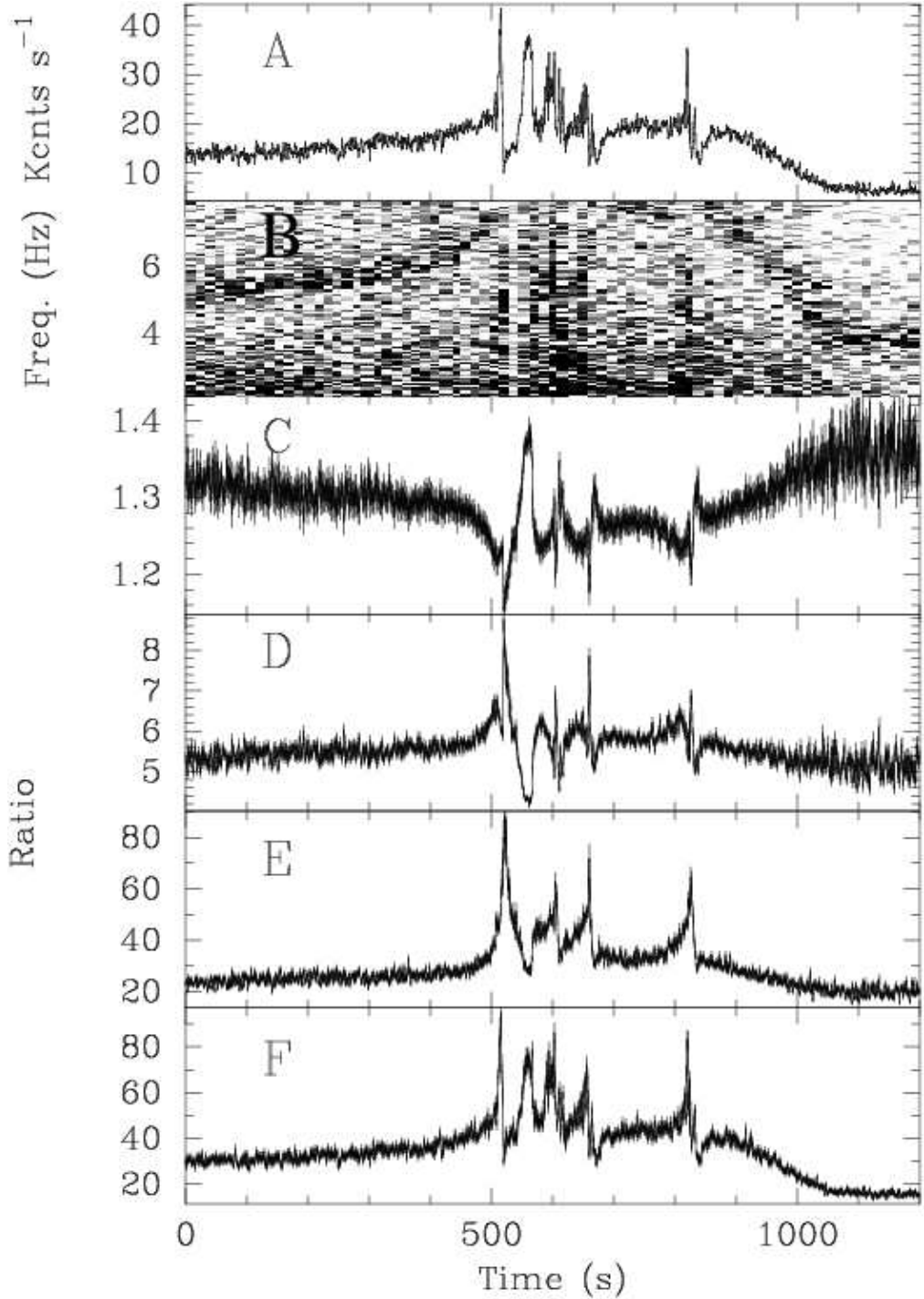}
\caption{RXTE-PCA observation of the GRS~1915+105 on 2001 July 16 (ObsId:
60702-01-01-03). Panel A and B shows the 2--60 keV light curve and the
dynamic power density spectra, respectively. Panels C to F show the ratio 
of the light curve in the 2--60 keV energy range to that in 2--8, 8--15, 
15--23, and 23--60 keV energy ranges, respectively. Absence of QPO during 
the dips (Panel B) and the inverted shape of dips in panels D and E suggest 
that the Comptonized spectral component is absent during the dips.}
\label{jul16_rat}
\end{figure*}

The presence of short duration and irregular flares and dips in the X-ray
light curves of RXTE observations on July 16 makes spectral fitting 
extremely difficult during the individual flare/dip events. To examine 
the properties of the source during these flares and dips, we follow the 
ratio method used by Vadawale et al. (2001a). Figure~\ref{jul16_rat} shows 
the Standard-1 light curve in the 2--60 keV energy band and the dynamic 
PDS along with the ratios of the Standard-1 light curve  to the light 
curves in 2--8, 8--15, 15--23 and 23--60 keV energy ranges. The second 
panel of the figure shows that the low frequency 4--7 Hz QPO, present 
in the steady part of the light curve, is absent during the dips. The 
third panel of the figure shows that during the dips, the decrease in the 
count-rate  in 2-8 keV range is less than that in the total count-rate. 
Inverted shapes of dips in the fourth and the fifth panels show that 
the counts in the energy band of 8-23 keV are preferentially decreased 
compared to the decrease in the total count rate (see Figures~\ref{beta} 
and \ref{theta} for a clear picture where the dips are of longer duration). 
Preferential decrease in the count rate in the 8-23 keV energy band 
strongly supports the inference drawn by Vadawale et al. (2001a) that 
the Comptonized component, which is dominant in this energy band, is 
absent during these dips. The absence of QPO during these dips 
(Figure~\ref{jul16_rat}, second panel), and the observation that the 
Comptonized component is responsible for the QPO in the hard ($C_{RL}$) 
states (Rao et al. 2000), further strengthen the suggestion that the 
Comptonized component vanishes during the dips. 

\section{Large flares observed by GBI and RXTE-ASM}

Our detailed analysis around the 2001 July radio flare (during which
a separate ejecta was detected by VLBI, exhibiting superluminal ejection)
shows that before the flare, the source was in $C_{RL}$ state consisting 
of three spectral components: a multi-color disk blackbody, a power-law, and
a Comptonized component. During the radio flare, the source showed a dipping 
behavior during which the Comptonized component of the X-ray spectra appears 
to be vanishing. A similar behavior of the source was observed during the 
superluminal ejection episode of 1999 June (Dhawan et al. 2003) and 
1997 October (Fender et al. 1999). The source was in $C_{RL}$ state before 
the flare (Rao et al. 2000; Vadawale et al. 2001b) and pointed observation 
during the flare showed soft dips (state A period) during which the 
Comptonized component vanishes. This leads to an interesting suggestion 
that the preceding $C_{RL}$ state and the presence of soft dips (state A 
periods) are the characteristics of all superluminal ejection episodes. 

\begin{figure*}
\vskip 18.5cm
\includegraphics{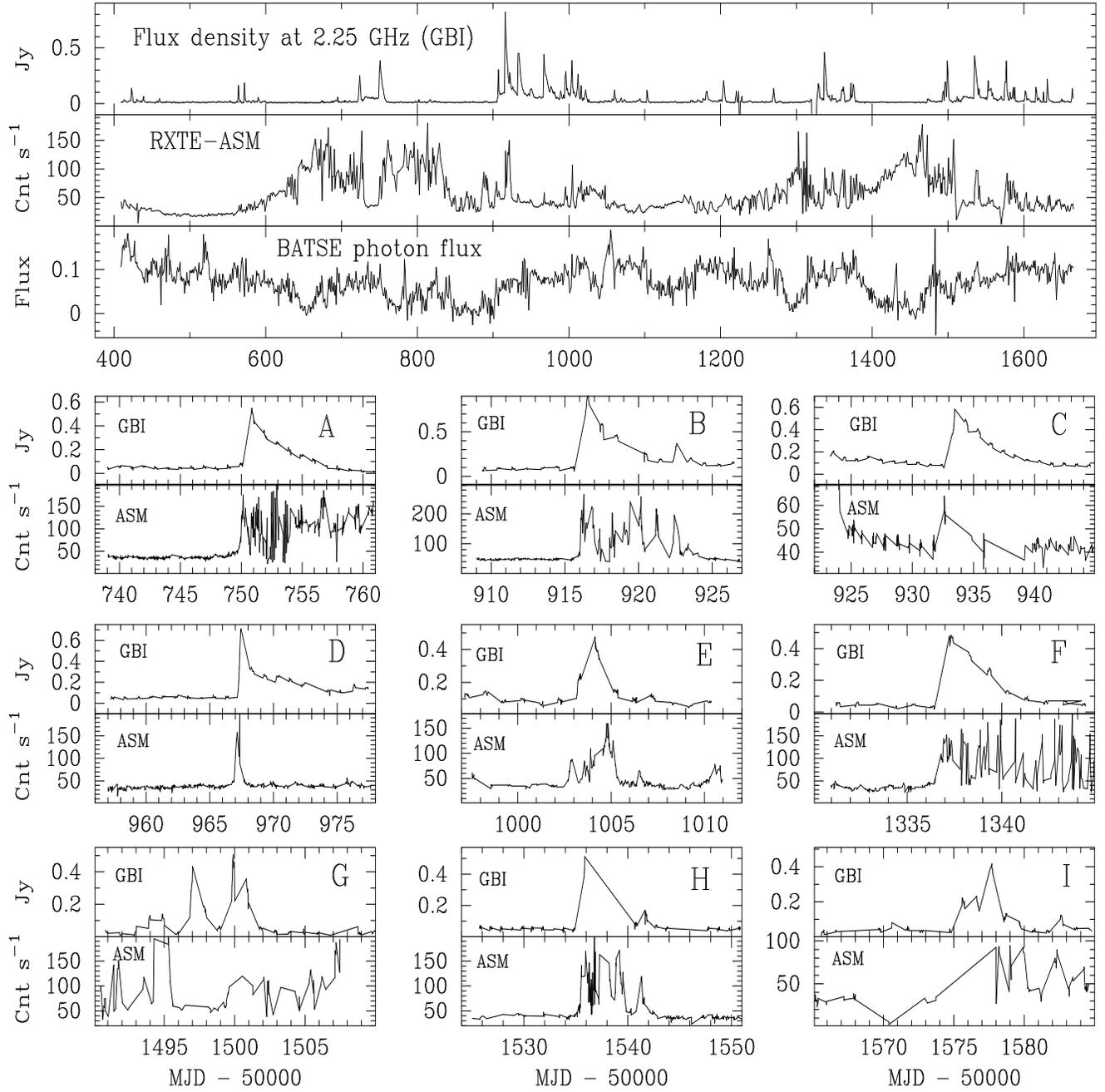}
\caption{Top three panels show GBI, RXTE/ASM, and CGRO/BATSE monitoring 
light curves of GRS~1915$+$105 from 1996 November to 2000 April. GBI and 
RXTE/ASM light curves during individual large flares, listed in 
Table~\ref{radio_flares}, are shown in panels A--I. The flares, shown in 
panels E, G and I, do not follow the fast rise and exponential decay (FRED) 
characteristic profile and hence are not considered as candidates for 
superluminal radio flares.}
\label{all_flares}
\end{figure*}

	To verify this suggestion we carried out a similar analysis for 
all possible superluminal radio flares observed during the period of 1996 
December to 2000 April when GBI radio monitoring data are available. Apart 
from the superluminal episodes in 1997 October and 1999 June, there is one 
more occasion, 1998 May, during which a separated ejecta is reported 
(Dhawan et al. 2000). However, a complete GBI monitoring radio light curve 
from 1996 December to 2000 April shows many other large flares and it might 
be possible that some such ejection episodes are missed. Therefore, we 
selected all large flares ($>$ 350 mJy at 2.25 GHz) from the GBI monitoring 
data. It is found that there are 9 flares satisfying this criteria, the 
characteristic properties of which are given in Table~\ref{radio_flares}. 
The GBI and ASM monitoring radio light curves of the source during all 
these 9 selected radio flares are shown in Figure~\ref{all_flares} along 
with the complete monitoring light curves from GBI, ASM and BATSE. All 
the large radio flares, during which separate ejecta are reported so far, 
show a fast rise and exponential decay (FRED) with decay time constant of 
the order of 3 days. It can be seen, from Table~\ref{radio_flares} and 
Figure~\ref{all_flares}, that the characteristic properties of 3 other 
radio flares (1998 April 13, 1998 June 3 and 1999 December 23) are found 
to be similar to those 3 flares during which superluminal ejection is 
reported. Hence we assume that all flares with the FRED morphology are 
superluminal flare episodes and we give the integrated flare energy (in 
units of Jy.day) in Table 2. The GBI light curve  also shows that the 
source was in radio ``plateau'' state prior to the all six `assumed' 
superluminal flares whereas the ASM light curves during these periods 
show that the source was in low and steady X-ray state. The properties 
of the three radio flares without FRED such as shape of the radio flares, 
variability in ASM light curve prior to the flares, are different from 
those seen during the superluminal radio flares.

\begin{figure*}
\vskip 12.5cm
\includegraphics{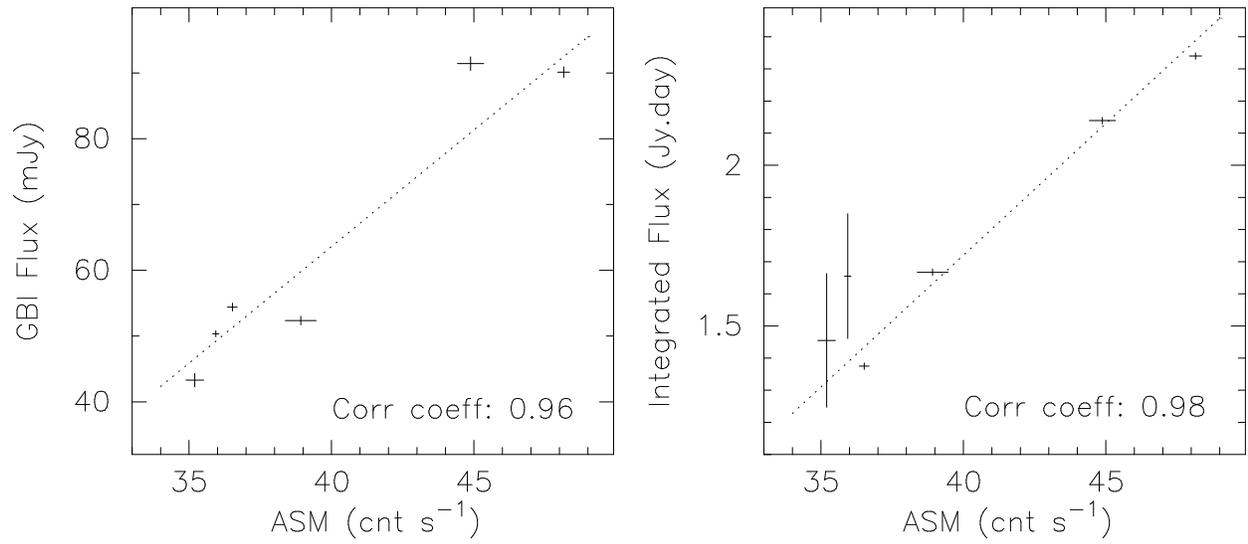}
\caption{Correlations between the average ASM count rate during the pre-flare
``plateau'' state and the average GBI flux during pre-flare ``plateau'' state
(left panel) as well as the integrated flux during the six selected flares 
(right panel). For six data points these correlations are significant at 
98 \% and 99 \% confidence level,  respectively.}
\label{corr}
\end{figure*}

A closer look at Table~\ref{radio_flares} reveals an interesting fact 
that the average ASM count rate and the GBI flux during the pre-flare 
``plateau'' state show a strong correlation suggesting a strong
coupling between the X-ray and radio emission.  In  Figure~\ref{corr} (left
panel) we show a plot of GBI flux against the ASM count rate along with
a best-fit straight line. We derive a  correlation coefficient of 0.96 
(98\% confidence level). This could be similar to the correlation between 
the X-ray and the radio emissions found in the Galactic black hole 
candidate sources (see eg. Gallo, Fender \& Pooley 2003; Choudhury et al. 
2003).

We also find that the average pre-flare ASM count rate and the total flux 
integrated over the duration of the flare are also strongly correlated. 
In Figure~\ref{corr} (right panel) we show this  correlation (with a 
correlation coefficient of 0.98) which is  significant at 99\% confidence 
level. The correlation between the pre-flare ASM count rate and the
integrated flare flux is particularly interesting, given the temporal
and spatial separation between the two types of emission. It shows that 
the pre-flare ``plateau'' state plays a crucial role in the formation
of the large superluminal radio flares.

\begin{figure*}
\vskip 18.9cm
\includegraphics{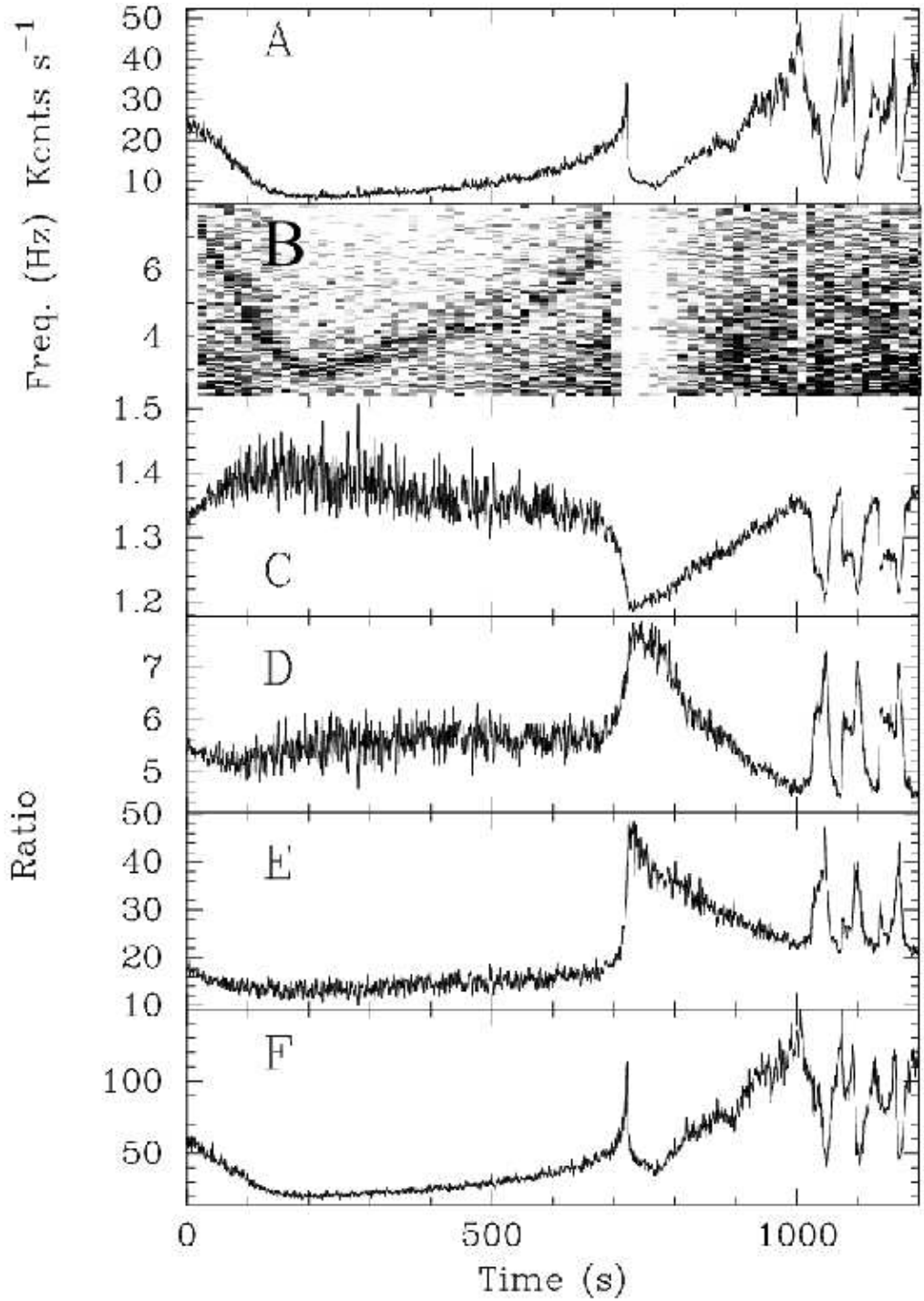}
\caption{RXTE-PCA observation of GRS~1915+105 on 1997 October 30 (ObsId:
20402-01-52-02). All panels are similar to those in Figure~\ref{jul16_rat}.}
\label{beta}
\end{figure*}

\begin{figure*}
\vskip 18.5cm
\includegraphics{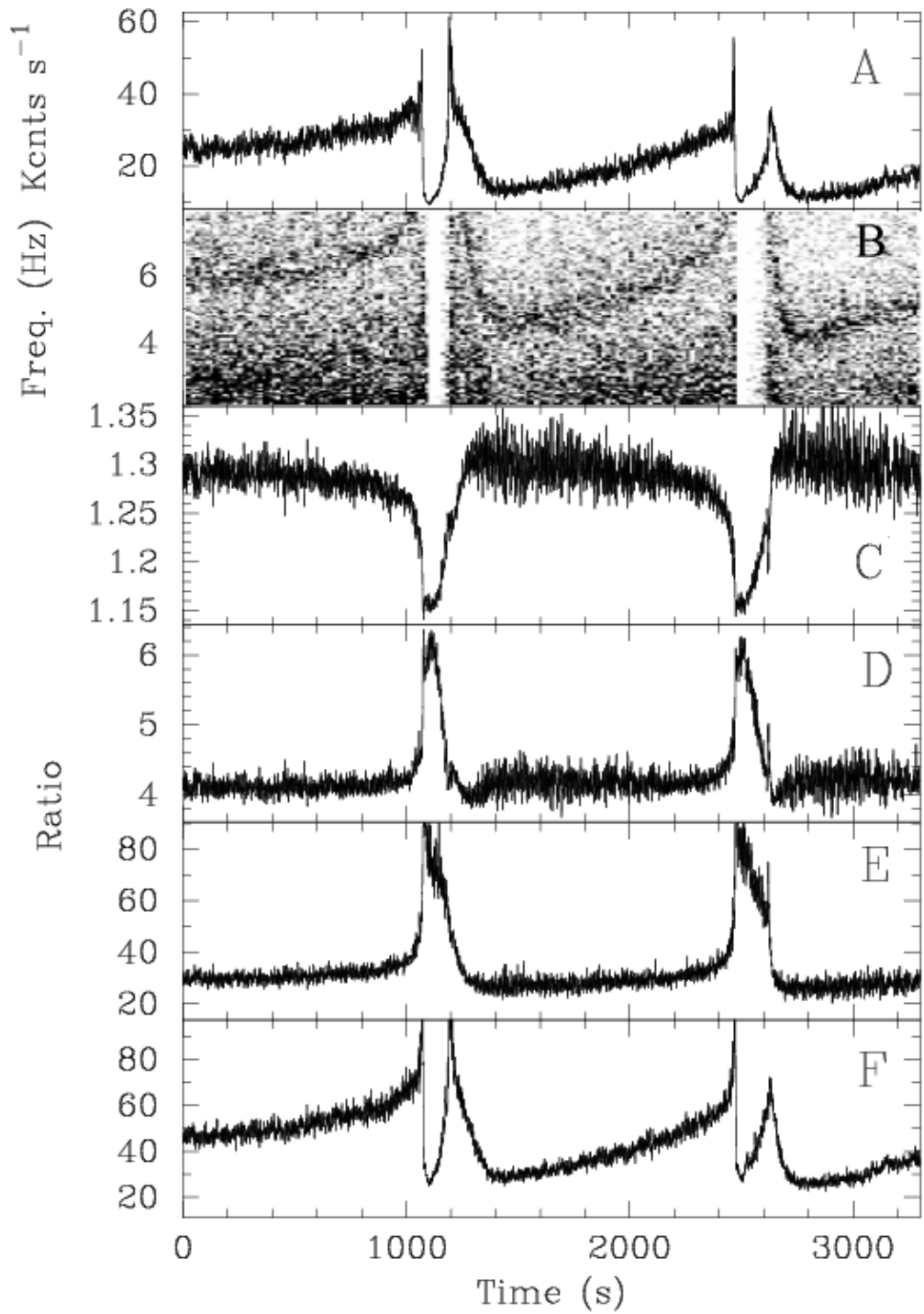}
\caption{RXTE-PCA observation of GRS~1915+105 on 1999 June 8 (ObsId:
40702-01-03-00). All panels are similar to those in Figure~\ref{jul16_rat}.}
\label{theta}
\end{figure*}

To quantify the timing and the spectral properties of the pre-flare states
of all the six selected flares, we have done a  detailed analysis of the 
RXTE pointed observations during these steady states. As the properties 
of the source are rather stable during these states, we analyzed two RXTE 
observations immediately before each flare. We also analyzed the RXTE 
observations on 1997 October 30 and 1999 June 8, which are the only 
pointed observations of the source during the rising phase of superluminal 
flares. We adopted a similar analysis procedure described in the previous 
section and fitted the same three-component model to the source spectra 
of all the observations. The log of observations and the best-fit spectral 
parameters are given in Table~\ref{all_year}. The results of analysis of 
the two pointed observations during the onset of superluminal flares are 
shown in Figure~\ref{beta} and \ref{theta}, which are similar to 
Figure~\ref{jul16_rat}.

We find that X-ray spectrum during all the pointed RXTE observations 
immediately before the selected flares can be fitted by our model consisting 
of three components: a disk-blackbody, a Comptonized component and a 
power-law. This confirms our three-component description of the $C_{RL}$ 
state. Both the available observations close to but after the onset of the 
radio flare shows dipping behavior as shown in Figures~\ref{beta} and 
\ref{theta}. It can be seen that, though the morphology of the X-ray light 
curves (classes according to Belloni et al. 2000) is different in all the 
pointed observations during the superluminal flares (Figures~\ref{jul16_rat}, 
\ref{beta} and \ref{theta}), the properties of the source during the dips 
are the same. The absence of QPOs during the dips in X-ray light curves 
(panel B of the above three figures) and the preferential decrease in 
8 -- 23 keV count rate (as seen from the inverted shape of the ratios in 
panels D and E of these figures) confirm the absence of the Comptonized 
component during the dips. We have verified, for a large number of 
observations containing similar soft dips (state A periods), that the 
characteristics of all the soft dips are similar. Thus it seems that 
the preferential decrease in $\sim$ 8 -- 25 keV range, suggesting the 
absence of the Comptonized component, is rather a general feature of all 
the soft dips (short state A periods) seen in this source. This leads to 
an interpretation that the central Compton cloud is ejected away during all
the soft dips.  Similar conclusion of mass ejection during dips of type A 
is drawn by Yadav (2001) on the basis of their spectral study during the 
class $\beta$.

\section{Discussion}

We have presented a detailed study of all the candidate superluminal 
ejection episodes of  GRS~1915+105, identified from the GBI, ASM, and 
BATSE monitoring observations. It should be noted here that there are at 
least two other episodes of large flares, first one during 2000 July-August, 
reported by Fender et al. (2002a) which follows the ``plateau'' state -- 
large flare sequence and the other during 2001 March reported in Fender 
et al. (2002b) for which the superluminal ejection was observed directly. 
We have verified that the results presented here are consistent with the 
RXTE pointed observations during 2000 July-August 2000 flare episode. 
For the 2001 March flare episode, unfortunately, both RXTE-PCA and 
RXTE-ASM coverage are poor and hence the pre-flare state cannot be 
established, but we believe that it must be $C_{RL}$.

Results presented in the previous sections confirm the suggestions that (i) 
the wide band X-ray spectrum during the $C_{RL}$ state has three spectral 
components: a multicolor disk-blackbody, a Comptonized component and a
power-law (ii) the soft dips (short periods of state A) are present in the 
X-ray light curve observed during the onset of superluminal flares, 
(iii) the Comptonized component is absent during the soft dips. These
results  lead to an interesting suggestion that the Compton cloud is 
ejected away during the soft dips. Formation of a central hot region can 
be explained under the Two Component Accretion Flow (TCAF) paradigm as 
the hot, quasi-spherical, post shock region near the black hole
(Chakrabarti \& Titarchuk 1995, Chakrabarti 2001 and references there in)
which is responsible for the presence of the Comptonized component (Rao 
et al. 2000). Ejection of the central Compton cloud can be explained if 
magnetic field is included in the TCAF framework (Nandi et al. 2001). 
The pre-flare $C_{RL}$ states are known to have a steady jet and the 
sudden ejection of matter in the previously established jet raises an 
interesting possibility for the origin of the superluminal flares in 
terms of the internal shock model.

\subsection{A Mechanism for superluminal motion?}

Apparent superluminal motion of blobs during the large radio flares of 
Galactic micro-quasars is one of the most spectacular phenomena of high 
energy astrophysics. GRS~1915+105 is the only source which shows repeated 
episodes of large radio flares exhibiting superluminal motion. The apparent 
superluminal motion can be understood as geometric effect due to very high
intrinsic velocity (i.e. close to $c$) of the moving blob at some angle 
with the line of sight. So far attempts have been made to understand the 
superluminal motions based on two different ideas: (1) matter blob moving 
at very high velocity and (2) shocks moving through a relatively slowly 
moving medium. Atoyan \& Aharonian (1999) showed that it is possible to 
reproduce the observed radio light curves of the separating blobs in 
terms of ejection and evolution of discrete plasmons, however, the 
instantaneous power requirement is very high in this model. Further they 
found that a single population of relativistic particles accelerated at 
the time of ejection of the plasmon cannot explain the observations and 
a continuous replenishment of relativistic particles is necessary. The 
continuous acceleration of particles comes naturally in the alternative 
model, the internal shock model proposed by Kaiser, Sunyaev \& Spruit 
(2000). They show that if the moving blobs are the traveling shock front 
arising due to differential velocity of the moving continuous jet material 
then it is possible to reproduce the observed light curves. The total 
energy requirement is the same as the plasmon model, however, the rate 
at which the energy supplied from the central source is much lower than 
that in the plasmon model.

The presence of the underlying continuous jet and the differential 
velocity of the moving matter are assumed {\em a priori} in the internal 
shock model and no attempt has been made to explain them. However, it can 
be seen that, in the context of the present work, these requirements are 
met quite naturally if the sudden ejection of the Compton cloud (i.e. the 
state A dip) occurs after a prolonged period of $C_{RL}$ state. The 
velocity of the ejected matter blob can be higher than the velocity of 
the matter in the underlying continuous jet, and hence as the ejected 
matter blob passes through the jet it creates an internal shock at the 
leading front where particles can be accelerated. Gallo et al. (2003) 
showed that, in other GBHCs, the velocity of the matter in steady jets 
is lower than that during transient events. This gives further confidence 
to the picture of the ejected mass driving through slower steady flow 
causing the internal shock. This moving internal shock can then be 
observed as separating blobs in the radio images. This suggestion 
provides a rather complete picture for the different types of radio 
emission from this source as described in the next section.

\subsection{The suggested scenario}\label{scenario}

The results obtained from our analysis as well as the various results 
on radio flares reported in the literature, can be summarized as follows.

\begin{itemize}

\item All superluminal radio flares, during which a separated ejecta 
is reported, as well as candidate superluminal flares for which there
is no report of observation, are preceded by the ``radio-loud hard 
($C_{RL}$) state''. 

\item The source spectrum during $C_{RL}$ state consists of three 
components: a multicolor disk-blackbody, a Comptonized component and 
a power-law. Large values of the derived inner disk radius suggest 
a geometry in which there is a hot spherical Compton cloud (responsible 
for the Comptonized component) surrounded by a thin disk (responsible 
for the disk-blackbody component).

\item Whenever there is a pointed X-ray observation during the onset of a
superluminal radio flare (3 out of 7 cases), the light curve shows soft 
dips (small state A episodes). During such dips, the Comptonized component 
is absent in the source spectrum.

\item High radio emission is observed from the source only during the
X-ray variability classes having soft dips (short state A periods) apart
from the radio ``plateau'' periods.

\item All superluminal flares show a characteristic fast rise and 
exponential decay profile in their radio light curve, which cannot be 
explained by the blob-ejection mechanism. Such profiles can be explained 
by the internal shock model of Kaiser et al.(1999). According to this 
model, the motion of the matter blob through the already present matter 
is seen as the moving ejecta during high spatial resolution radio 
observations.

\item The radio ``Plateau'' ($C_{RL}$) state is known to have steady jets
which eject a continuous stream of matter. Thus if a blob-ejection occurs
after a ``plateau'' state, it will satisfy the requirement of the internal
shock model and gives rise to a superluminal radio flare.

\item The total integrated flux during the superluminal flares is highly 
correlated with the observed X-ray flux during the preceding $C_{RL}$ states.
\end{itemize}

These points suggest a scenario for different types of radio emission
observed from this source.

\begin{itemize}
\item The $C_{RL}$ states have continuous jets and prolonged episodes 
of such states give rise to steady high radio emission with flat
spectrum which arises from the continuous jet. This is a fairly robust
statement as the compact jets have actually been observed during the
episodes of $C_{RL}$ states (Dhawan et al. 2000).

\item The ejection of the Compton cloud during a soft (state A) dip
gives rise to a small radio flare. Repeated occurrence of such dips 
gives rise to radio oscillation events. It should be noted that the 
amplitude of the isolated flares or the radio oscillation events will 
depend on various aspects like the amount of mass outflow during the
immediately previous $C_{RL}$ state, the amount of mass ejected during
the dips etc. Hence, it might not be possible to always observe small
radio flares or oscillation when such dips are present in X-ray, though, 
in such cases the average radio emission should be high. However, this 
means that whenever radio oscillations are observed, the soft dips must 
be present in the X-ray light curves.

\item If the occurrence of the soft (state A) dip is preceded by a 
prolonged period of $C_{RL}$ state having a continuous jet then
it satisfies the requirements of the internal shock model and gives
rise to the observed superluminal motion of blobs as seen in high 
resolution radio images.
\end{itemize}

\begin{figure*}
\centering
\vskip 16.5cm
\includegraphics{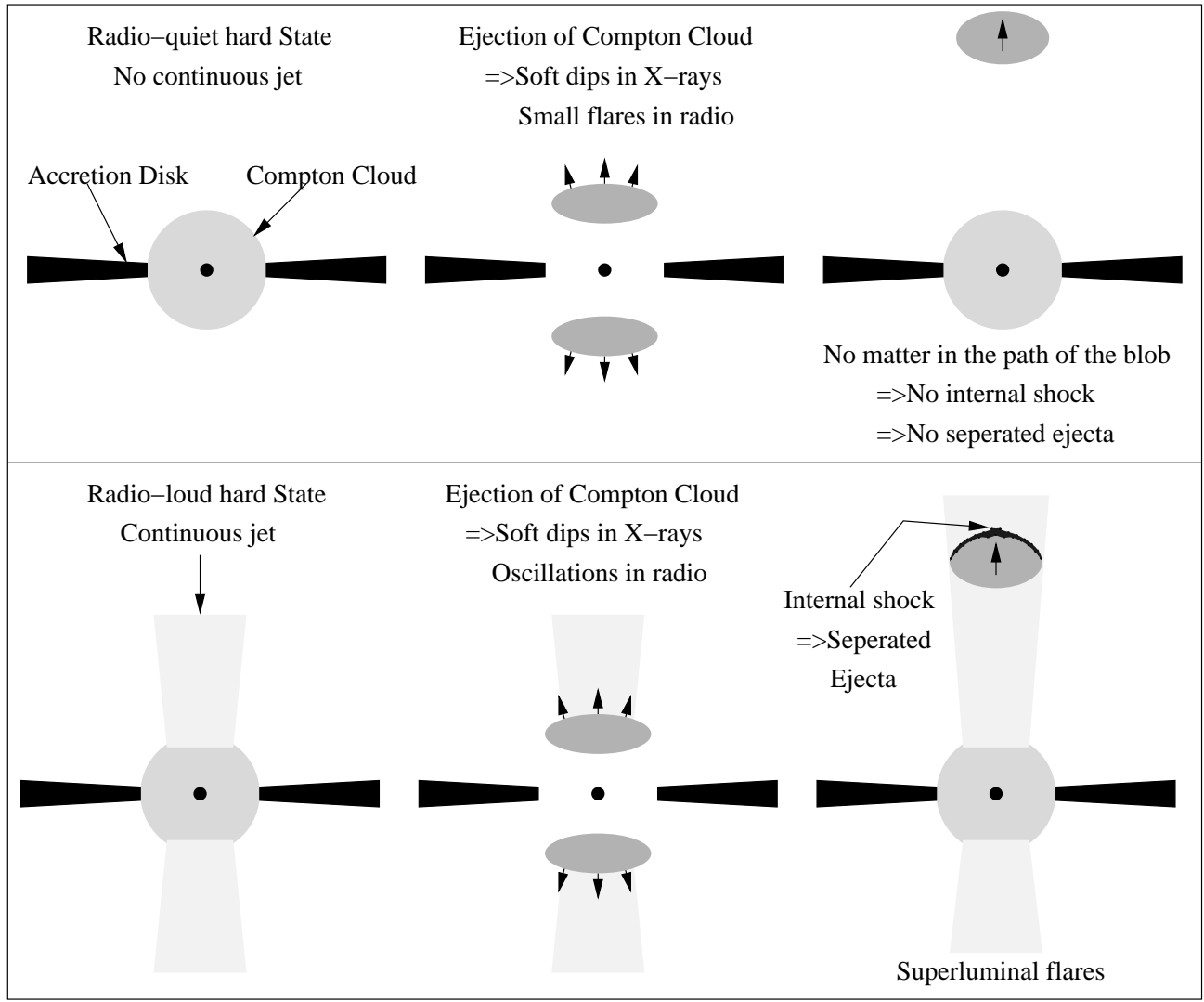}
\caption{A schematic view of the proposed scenario. If there is no continuous
matter outflow in the hard state then it is a radio-quiet hard state. The 
ejection of the Compton cloud, in such a case, results in an isolated flare 
(or oscillations if the ejection is repeated). If there is a continuous 
matter outflow in the hard state then it is a radio-loud hard state. The 
ejection of the Compton cloud, in such a  case, results in the separated 
ejecta showing superluminal motion.}
\label{schematic}
\end{figure*}

A schematic picture of this scenario is shown in Figure~\ref{schematic}. 
According to this scenario, any short time variation in the radio flux is 
caused by the occurrence of the soft (state A) dips during which the Compton 
cloud is ejected. This is slightly different from the inference drawn by 
Klein-Wolt et al. (2002) that only state C episodes are related to the 
radio emission. They find that it is necessary to have the state C episodes 
well-separated in order to have radio oscillation events. However, there 
are X-ray variability classes like $\lambda$ which show long, well 
separated periods of state C, but the average radio flux during class 
$\lambda$ is much lower compared to the classes $\beta$, $\theta$ 
which also have well separated long state C periods (Naik \& Rao 2000). 
This can be explained in our scenario according to which, the state C 
periods in class $\lambda$ are actually state $C_{RQ}$ periods whereas the 
state C periods in classes $\beta$, $\theta$ are actually state $C_{RL}$ 
periods. Further, it can be seen that such well-separated state C episodes, 
when the radio oscillations are observed, are almost always accompanied by 
state A dips. Thus the one-to-one relation between the state C episodes 
and the radio oscillation events found by Klein-Wolt et al. (2002) is 
valid in our scenario also. It should be noted that Figure 3g of 
Klein-Wolt et al. (2002) does not show any oscillations in the presence 
of class $\theta$. According to our scenario it is possible not to 
have radio oscillations in the presence of state A dips, i.e. because 
of close ejection events etc., but not vise-versa. As a counter example,  
Figure 10 of Dhawan et al. (2000) show clear oscillations during the 
presence of class $\theta$. This picture is further supported by the 
pointed radio observation near the onset of the superluminal flare in 
1997 October by Fender et. al (1999), which shows core oscillations 
as well as superluminal jets on 1997  October 30 and 31. On both these 
days the source was in class $\beta$ which shows A type dips after C state 
(Yadav 2001). The hypothesis of mass ejection during the soft dips is also 
supported by the calculation of time delay between emissions at different 
frequencies as presented by Ishwara-Chandra et al. (2002). They show that 
the IR emission follows the mass ejection with almost zero time delay. It 
is known that IR flares start at the time of spike (in beta class) which 
marks the beginning of A type dips (Eikenberry et al. 1998).

There are some important cross-checks for this scenario. First of all, it 
predicts that the separated ejecta should not be observed for any radio 
flare other than those having FRED profile and are preceded by $C_{RL}$ 
states. Also whenever radio oscillations are observed, the soft (state A) 
dips must be present in the X-ray light curve. A large data base of radio 
and X-ray observation of GRS~1915+105 has been collected so far and all 
the reported observations satisfies these criteria. However, the most 
important aspect of this scenario, to be tested with future observations, 
is the three-component description of the wide band X-ray spectra during 
the $C_{RL}$ state. This can be verified, for example, by high sensitivity 
observations above 50 keV, where the differences in the variation of the 
RMS power of the QPO with energy between $C_{RL}$ and $C_{RQ}$ state will 
show the existence of additional component which is not participating in 
the QPO. It should be noted that according to this picture, the low 
luminosity and soft spectrum of state A is due to the absence of the 
central Compton cloud. However this applies only to the short duration 
state A period appearing as soft dips. There exists a variability class 
$\phi$ during which the source appears to be in state A for much longer 
duration ($\sim$days) which can not be explained in the present picture
(see, however, Vadawale et al. 2003 for an attempt to understand the
class $\phi$ as due to obscuration of the Compton cloud).

	This mechanism suggests that whenever there is a $C_{RL}$ state,
the next soft X-ray dip events will give rise to a superluminal flare. 
Thus it provides a very important handle in predicting the occurrence 
of a flare which can help to plan new coordinated observation of large 
flares. Such observations can lead to further understanding of very 
important questions like exact mechanism of ejection, cause of ejection,
acceleration of the matter etc. It can be seen that  a  continuous 
hard X-ray and radio monitoring is very important in distinguishing 
$C_{RQ}$ state and $C_{RL}$ states. Unfortunately, after CGRO-BATSE 
there is no hard X-ray monitor currently available making an important 
observation handle on the occurrence of $C_{RL}$ state unavailable. 
The Solar X-ray Spectrometer (SOXS), which is a small phoswich type 
open hard X-ray detector (Malkar et al. 2003), will also perform  daily 
monitoring of a few bright sources and hence might be useful in predicting 
large superluminal flares from this source.

\section{Conclusion}

	In this paper we have tried to explain the observed association
between the radio ``plateau'' or the $C_{RL}$ states and the superluminal 
flares in GRS~1915+105. All $C_{RL}$ states have very similar X-ray 
spectral properties and show a three-component spectra. We have further 
shown that the soft X-ray dips are observed near the onset of radio 
flares. During such dips the Comptonized component is absent, which 
we propose as due to the ejection of the central Compton cloud. This 
ejected matter is then responsible for the low frequency emission
which can be observed as an isolated flare or regular oscillations (if
the dips are repeating). Such ejections might give rise to superluminal
flares if these dip events are preceded by a $C_{RL}$ state according to 
the internal shock model for the superluminal flares. Two definite 
predictions of this scenario are that (1) whenever radio oscillations 
are observed, the soft dips will be present in the X-ray light curves 
and (2) if $C_{RL}$ is not preceded by any radio flare it will not have 
the characteristic FRED shape and will not show separated ejecta. 
It also  suggests that in the future whenever $C_{RL}$ is observed, 
there are very good chances to catch a superluminal flare in `live 
action' and well coordinated multi wavelength observations of such an 
event will help understanding of the fundamental questions underlying 
such events.

\section{Acknowledgments}

We thank the members of the RXTE, BATSE, and GBI teams for making the data 
publicly available. This research has made extensive use of data obtained 
through the High Energy Astrophysics Science Archive Research Center 
(HEASARC) online service, provided by NASA/Goddard Space Flight Center. 
The Green Bank Interferometer (GBI) is a facility of the National Science 
Foundation operated by the NRAO in support of NASA High Energy Astrophysics
programs.

\clearpage

\begin{table}[t]
\caption{Results of analysis of pointed RXTE observation during and prior
to the large radio flare observed on 2001 July 16.}
\begin{center}
\scriptsize
\begin{tabular}{cccccccc}
\multicolumn{8}{c}{{\bf A:} Observation log, QPO frequency and the observed 
fluxes} \\
\hline
\hline
ObsId$^1$ & Date & Start     &$\nu_{QPO}$ &\multicolumn{4}{c}{3 -- 150 keV 
flux (in 10$^{-8}$ erg cm$^{-2}$ s$^{-1}$)}\\
  &  & Time (UT) & (Hz) &F$_{total}$ & F$_{dbb}$ & F$_{ctt}$ & F$_{pow}$\\
\hline
A-65-00 &June 30 &21:35:28  &3.124 &2.44 &0.27 &1.38 &0.77\\
A-66-00 &July 06 &21:18:24  &1.901 &1.95 &0.20 &0.95 &0.71\\
A-66-01 &July 07 &00:41:36  &1.908 &1.92 &0.22 &1.01 &0.60\\
B-04-00 &July 09 &01:59:28  &2.202 &2.07 &0.23 &1.09 &0.64\\
B-04-05 &July 10 &05:52:16  &2.738 &2.33 &0.29 &0.71 &1.23\\
B-04-08 &July 11 &01:41:36  &2.434 &2.09 &0.24 &0.75 &1.01\\
\\
C-01-02 &July 16 &00:51:28  &4.341 &3.65 &0.46 &0.52 &2.59\\
C-01-01 &July 16 &02:30:40  &3.998 &3.25 &0.50 &1.52 &1.07\\
C-01-00 &July 16 &08:10:56  &4.533 &3.70 &0.63 &2.28 &0.61\\
C-01-03 &July 16 &11:20:48  &4.194 &3.07 &0.53 &1.42 &0.95\\
C-01-04 &July 16 &18:03:28  &4.874 &3.90 &0.68 &1.47 &1.55\\
C-01-05 &July 16 &21:21:20  &4.705 &3.72 &0.62 &1.38 &1.55\\
C-01-06 &July 16 &23:02:40  &4.895 &3.89 &0.67 &1.04 &1.99\\
\hline
\\
\multicolumn{8}{c}{{\bf B:} Best fit Spectral parameters$^2$} \\
\hline
\hline
ObsId$^1$ & N$_H$ & kT$_{in}$  & kT$_e$ & $\tau$ & $\Gamma$ &R$_{in}$ $^3$ 
&$\chi^2_{\nu}$(dof) \\
 &(10$^{22}$ cm$^{-2}$) &(keV) &(keV) &    &     &(km)    &\\  
\hline
A-65-00 & 1.65$^{+0.13}_{-0.16}$ &1.06$^{+0.09}_{-0.02}$  & 
7.95$^{+0.07}_{-0.14}$ & 6.38$^{+0.06}_{-0.05}$ & 2.17$^{+0.01}_{-0.01}$ &
34.79 &1.98(87) \\
A-66-00 & 13.69$^{+0.54}_{-0.88}$ & 0.53$^{+0.01}_{-0.01}$ & 
7.97$^{+1.16}_{-0.59}$ & 6.88$^{+0.59}_{-0.64}$ & 2.27$^{+0.17}_{-0.23}$ &
558.98 &0.97(87) \\
A-66-01 & 14.22$^{+0.88}_{-0.85}$ & 0.54$^{+0.01}_{-0.01}$ & 
7.32$^{+0.74}_{-0.98}$ & 7.26$^{+1.82}_{-0.65}$ & 2.17$^{+0.26}_{-0.33}$ &
567.59  & 0.98(81)\\
B-04-00 & 14.53$^{+0.95}_{-0.89}$ & 0.52$^{+0.01}_{-0.01}$ & 
9.77$^{+1.45}_{-1.12}$ & 6.01$^{+0.68}_{-0.84}$ & 2.35$^{+0.09}_{-0.11}$ &
721.83  & 1.40(87)\\
B-04-05 & 14.73$^{+0.59}_{-0.86}$ & 0.52$^{+0.01}_{-0.01}$ & 
6.73$^{+0.83}_{-0.85}$ & 8.08$^{+1.56}_{-0.82}$ & 2.52$^{+0.08}_{-0.17}$ &
453.99  & 0.93(87)\\
B-04-08 & 13.10$^{+0.67}_{-0.99}$ & 0.54$^{+0.01}_{-0.01}$ & 
7.02$^{+1.42}_{-0.81}$ & 7.83$^{+0.88}_{-1.06}$ & 2.46$^{+0.07}_{-0.27}$ &
575.62  & 0.98(87)\\
\\
C-01-02 & 14.46$^{+0.54}_{-0.34}$ & 0.53$^{+0.01}_{-0.01}$ & 
7.88$^{+1.22}_{-1.38}$ & 8.09$^{+4.59}_{-1.41}$ & 2.83$^{+0.04}_{-0.06}$ &
914.69  & 1.04(87)\\
C-01-01 & 13.47$^{+0.56}_{-0.65}$ & 0.56$^{+0.01}_{-0.01}$ & 
8.85$^{+1.96}_{-1.44}$ & 5.31$^{+0.92}_{-0.53}$ & 2.52$^{+0.13}_{-0.47}$ &
728.57   & 1.23(87)\\
C-01-00 & 14.46$^{+0.45}_{-0.40}$ & 0.56$^{+0.01}_{-0.01}$ & 
10.86$^{+0.35}_{-1.26}$ & 4.15$^{+0.09}_{-0.11}$ & 2.31$^{+0.46}_{-0.38}$ &
847.52  & 0.99(87) \\
C-01-03 & 13.90$^{+0.58}_{-0.58}$ & 0.57$^{+0.01}_{-0.01}$ & 
9.21$^{+0.99}_{-0.92}$ & 5.16$^{+0.72}_{-0.50}$ & 2.53$^{+0.14}_{-0.31}$ &
724.04  & 1.16(87)\\
C-01-04 & 14.41$^{+0.48}_{-0.48}$ & 0.57$^{+0.01}_{-0.01}$ & 
8.35$^{+0.28}_{-0.23}$ & 5.18$^{+0.12}_{-0.29}$ & 2.58$^{+0.06}_{-0.16}$ &
830.45  & 0.93(87)\\
C-01-05 & 12.97$^{+0.61}_{-0.59}$ & 0.59$^{+0.01}_{-0.01}$ & 
8.23$^{+1.45}_{-2.88}$ & 5.39$^{+0.67}_{-0.69}$ & 2.56$^{+0.44}_{-0.39}$ &
605.30  & 0.68(87)\\
C-01-06 & 14.39$^{+0.43}_{-0.56}$ & 0.56$^{+0.01}_{-0.01}$ & 
7.12$^{+0.36}_{-0.42}$ & 6.29$^{+0.30}_{-0.38}$ & 2.71$^{+0.05}_{-0.14}$ &
835.85   & 1.02(87)\\
\hline
\hline
\multicolumn{8}{l}{{\bf $^1$}: A = 50703-01, B = 60405-01, C = 60702-01}\\
\multicolumn{8}{l}{{\bf $^2$}:  Model consists of a disk-blackbody, CompTT 
and power-law (see text)}\\
\multicolumn{8}{l}{{\bf $^3$}: $R_{in}$ derived assuming distance $D=12.5$ 
kpc and inclination angle, $\theta=70^{o}$}\\
\end{tabular}
\label{jul}
\end{center}
\end{table}

\begin{table}[t]
\caption{Properties of all radio flares stronger than 350 mJy at 2.25 GHz
present in the GBI monitoring light curve.}
\begin{center}
\scriptsize
\begin{tabular}{|ccccc|ccccc|}
\hline
\hline
\multicolumn{5}{|c|}{Radio flare properties }  &\multicolumn{5}{|c|}{Pre-flare ``plateau'' properties} \\
\hline 
MJD &Date & Peak flux  &Decay Time &Integrated$^a$ &Duration &Ave. GBI &
Ave. ASM  &ASM   &No. of  \\
 & &(mJy) &(day) &flux (Jy.day) &(day) &flux (mJy) &count s$^{-1}$ &
Variability (\%) &PCA Obs. \\
\hline
50751 &1997 Oct 31  &550 &3.16    &1.65$^c$ &19 &50.3  &35.9 &5.44  &5 \\
50916 &1998 Apr 13  &920 &3.01    &2.34 &6  &90.1  &48.2 &2.70  &4 \\
50933 &1998 Apr 30  &580 &3.82    &2.14 &6  &91.4 &44.9 &3.31  &2 \\
50967 &1998 Jun 03  &710 &2.13    &1.37 &12 &54.4  &36.5 &5.10  &12 \\
51004 &1998 Jul 10  &480 &No FRED &--   &7  &112.7 &43.6 &18.82 &--\\
51337 &1999 Jun 08  &490 &2.68    &1.46$^c$ &5  &43.3  &35.2 &6.05  &7  \\
51499 &1999 Nov 17  &510 &No FRED &--   &8  &82.8  &69.2 &16.60 &--\\
51535 &1999 Dec 23  &510 &3.86$^b$&1.67 &10 &52.3  &38.9 &3.96  &7  \\
51577 &2000 Feb 03  &420 &No FRED &--   &12 &69.8  &30.0 &11.50 &4\\
\hline
\hline
\multicolumn{10}{l}{{\bf a:} Integrated flux is obtained by fitting an 
exponential function to the flare profile and integrating the function }\\
\multicolumn{10}{l}{~~~~over duration of 3$\times$decay constant. Typical 
errors are $\pm$0.01, unless stated} \\
\multicolumn{10}{l}{{\bf b:} This is uncertain because of large data gaps 
in the GBI monitoring data} \\
\multicolumn{10}{l}{{\bf c:} Typical errors are $\pm$0.2 due to uncertain 
start time. }\\
\end{tabular}
\label{radio_flares}
\end{center}
\end{table}

\begin{table}[t]
\caption{Results of analysis of two pointed RXTE observations immediately 
prior to the radio flares selected from Table~\ref{radio_flares}.}
\begin{center}
\scriptsize
\begin{tabular}{cccccccc}
\multicolumn{8}{c}{{\bf A:} Observation log, QPO frequency and the observed 
fluxes } \\
\hline
\hline
ObsId$^1$ & Date & Start     &$\nu_{QPO}$ &\multicolumn{4}{c}{3 -- 150 keV 
flux (in 10$^{-8}$ ergs s$^{-1}$)}\\
 & & Time (UT) & (Hz) &F$_{total}$ & F$_{dbb}$ & F$_{ctt}$ & F$_{pow}$\\
\hline
\hline
P-51-00  &1997 Oct 22 &07:04:48 &1.401  &2.012  &0.175 &0.437 &1.301 \\
P-52-000 &1997 Oct 25 &06:30:40 &1.882  &2.042  &0.245 &0.412 &1.258 \\ 
Q-09-01  &1998 Apr 10 &00:02:24 &2.175  &2.439  &0.313 &0.826 &1.134 \\
Q-10-00  &1998 Apr 11 &09:28:32 &1.739  &2.423  &0.291 &0.729 &1.247 \\
R-15-00  &1998 Apr 22 &21:16:00 &1.458  &2.443  &0.265 &1.029 &0.999 \\
R-16-00  &1998 Apr 28 &16:16:48 &1.399  &2.335  &0.288 &1.044 &0.828 \\
R-20-00  &1998 May 24 &19:27:28 &0.692  &2.131  &0.165 &0.876 &0.993 \\
R-21-00  &1998 May 31 &19:25:20 &1.763  &2.144  &0.252 &0.846 &0.912 \\
S-16-02  &1999 Jun 03 &03:02:24 &1.759  &1.978  &0.229 &0.727 &0.919 \\
S-17-01  &1999 Jun 07 &02:52:32 &2.874  &2.295  &0.313 &0.664 &1.182 \\
S-42-00  &1999 Dec 13 &11:56:32 &1.435  &2.137  &0.245 &0.680 &1.065 \\
S-43-00  &1999 Dec 21 &11:32:16 &2.121  &2.231  &0.253 &0.710 &1.158 \\
\hline
\hline
\multicolumn{8}{c}{{\bf B:} Best fit Spectral parameters$^2$} \\
\hline
\hline
ObsId$^1$ & N$_H$ & kT$_{in}$  & kT$_e$ & $\tau$ & $\Gamma$ &R$_{in}$ $^3$ &
$\chi^2_{\nu}$(dof) \\
 &(10$^{22}$ cm$^{-2}$) &(keV) &(keV) &    &    &(km)    &\\  
\hline
P-51-00 & 13.27$^{+0.66}_{-0.91}$ & 0.48$^{+0.01}_{-0.01}$ & 
4.33$^{+0.16}_{-0.23}$ & 11.90$^{+1.51}_{-1.74}$ & 2.75$^{+0.03}_{-0.01}$ & 
787.22 & 1.34(94) \\
P-52-000 & 14.68$^{+0.73}_{-0.68}$ & 0.48$^{+0.01}_{-0.01}$ & 
4.13$^{+0.30}_{-0.25}$ & 11.49$^{+0.78}_{-2.78}$ & 2.76$^{+0.10}_{-0.04}$ & 
1000.98 & 1.07(94) \\
Q-09-01 & 12.39$^{+0.84}_{-0.16}$ & 0.51$^{+0.01}_{-0.01}$ & 
3.92$^{+0.17}_{-0.21}$ & 9.00$^{+0.61}_{-0.39}$ & 2.79$^{+0.02}_{-0.10}$ & 
755.98 & 1.05(94)  \\
Q-10-00 & 14.95$^{+0.82}_{-0.65}$ & 0.46$^{+0.01}_{-0.01}$ & 
4.04$^{+0.13}_{-0.30}$ & 9.65$^{+0.86}_{-0.80}$ & 2.97$^{+0.09}_{-0.03}$ & 
1392.43 & 0.83(94)  \\
R-15-00 & 11.12$^{+0.85}_{-0.24}$ & 0.52$^{+0.01}_{-0.01}$ & 
4.20$^{+0.16}_{-0.07}$ & 9.07$^{+0.59}_{-0.08}$ & 2.72$^{+0.03}_{-0.02}$ & 
607.12 & 1.25(94)  \\
R-16-00 & 13.55$^{+0.78}_{-0.31}$ & 0.50$^{+0.01}_{-0.01}$ & 
4.18$^{+0.15}_{-0.22}$ & 8.75$^{+1.00}_{-0.44}$ & 2.50$^{+0.12}_{-0.19}$ &
 831.36 & 1.43(94)  \\
R-20-00 & 14.23$^{+0.35}_{-0.65}$ & 0.50$^{+0.01}_{-0.01}$ & 
5.20$^{+0.20}_{-0.12}$ & 8.66$^{+0.37}_{-0.72}$ & 2.61$^{+0.03}_{-0.05}$ & 
691.93 & 1.07(94)  \\
R-21-00 & 12.99$^{+0.81}_{-0.32}$ & 0.51$^{+0.01}_{-0.01}$ & 
4.96$^{+0.22}_{-0.51}$ & 8.46$^{+0.97}_{-0.68}$ & 2.69$^{+0.24}_{-0.06}$ & 
741.65 & 0.93(94)  \\
S-16-02 & 12.43$^{+0.95}_{-1.15}$ & 0.51$^{+0.01}_{-0.01}$ & 
5.98$^{+0.80}_{-0.60}$ & 7.68$^{+0.73}_{-1.29}$ & 2.52$^{+0.23}_{-0.12}$ & 
660.93 & 1.16(88)  \\
S-17-01 & 13.12$^{+0.44}_{-0.74}$ & 0.50$^{+0.01}_{-0.01}$ & 
5.58$^{+0.71}_{-0.38}$ & 7.90$^{+0.02}_{-1.27}$ & 2.75$^{+0.10}_{-0.06}$ & 
895.49 & 0.93(87) \\
S-42-00 & 12.94$^{+1.34}_{-0.51}$ & 0.51$^{+0.02}_{-0.01}$ & 
4.29$^{+0.24}_{-0.48}$ & 10.93$^{+2.19}_{-0.96}$ & 2.69$^{+0.13}_{-0.07}$ & 
710.50 & 1.00(88) \\
S-43-00 & 13.61$^{+1.17}_{-0.94}$ & 0.48$^{+0.02}_{-0.01}$ & 
5.25$^{+0.76}_{-0.39}$ & 8.29$^{+1.07}_{-2.23}$ & 2.82$^{+0.03}_{-0.07}$ & 
981.34 & 0.81(88) \\
\hline
\multicolumn{8}{l}{{\bf $^1$}: P = 20402-01, Q = 30402-01, R = 30703-01, 
S = 40703-01}\\
\multicolumn{8}{l}{{\bf $^2$}:  Model consists of a disk-blackbody, CompTT 
and power-law (see text)}\\
\multicolumn{8}{l}{{\bf $^3$}: $R_{in}$ derived assuming distance $D=12.5$ 
kpc and inclination angle, $\theta=70^{o}$}\\
\end{tabular}
\label{all_year}
\end{center}
\end{table}

\end{document}